\documentclass[letterpaper]{article} 
\usepackage{aaai2026}  
\usepackage{times}  
\usepackage{helvet}  
\usepackage{courier}  
\usepackage[hyphens]{url}  
\usepackage{graphicx} 

\usepackage{graphicx} 
\usepackage{tabularx}
\usepackage{multirow}
\usepackage{bm}

\usepackage{amssymb}
\usepackage{mathtools} 
\usepackage{stackengine}
\usepackage{tikz}
\usepackage{amsmath}
\usepackage{mathrsfs}
\usepackage{bm}

\usepackage{tabularx}

\usepackage{natbib} 
\usepackage{caption} 
\usepackage{graphics} 
\usepackage{epsfig} 
\usepackage{times} 
\usepackage{amsmath} 
\usepackage{amssymb}  
\usepackage{booktabs}
\usepackage{multirow}  
\usepackage{makecell}  
\usepackage{graphicx}
\usepackage{graphicx}
\usepackage{amsmath}
\usepackage{amssymb}
\usepackage{booktabs} 
\usepackage{multirow} 
\usepackage{url}
\usepackage{xcolor}
\usepackage{algorithm}
\usepackage[noend]{algpseudocode}  
\usepackage{cite}     
\usepackage{lipsum} 

\usepackage{scalerel}

\usepackage{caption}
\usepackage{booktabs}
\usepackage{subcaption}
\usepackage{adjustbox}
\usepackage{amsmath}

\usepackage{natbib}  
\usepackage{caption} 
\usepackage{newfloat}
\usepackage{listings}
\DeclareCaptionStyle{ruled}{labelfont=normalfont,labelsep=colon,strut=off} 
\floatstyle{ruled}
\newfloat{listing}{tb}{lst}{}
\floatname{listing}{Listing}
\title{From IDs to Semantics: A Generative Framework for Cross-Domain Recommendation with Adaptive Semantic Tokenization}

\author{
  Peiyu Hu\equalcontrib\textsuperscript{\rm 1,2},
  Wayne Lu\equalcontrib\textsuperscript{\rm 1,2},
  Jia Wang\textsuperscript{\rm 1,2}\thanks{Corresponding author.}
}
\affiliations{
  \textsuperscript{\rm 1}Xi'an Jiaotong--Liverpool University, Suzhou, China\\
  \textsuperscript{\rm 2}University of Liverpool, Liverpool, United Kingdom\\
  \{peiyuhu30, lu.wayne0603\}@gmail.com, Jia.Wang02@xjtlu.edu.cn
}

\usepackage{bibentry}

\begin{document}

\maketitle

\begin{abstract}

Cross-domain recommendation (CDR) is crucial for improving recommendation accuracy and generalization, yet traditional methods are often hindered by the reliance on shared user/item IDs, which are unavailable in most real-world scenarios. Consequently, many efforts have focused on learning disentangled representations through multi-domain joint training to bridge the domain gaps. Recent Large Language Model (LLM)-based approaches show promise, they still face critical challenges, including: (1) the \textbf{item ID tokenization dilemma}, which leads to vocabulary explosion and fails to capture high-order collaborative knowledge; and (2) \textbf{insufficient domain-specific modeling} for the complex evolution of user interests and item semantics. To address these limitations, we propose \textbf{GenCDR}, a novel \textbf{Gen}erative \textbf{C}ross-\textbf{D}omain \textbf{R}ecommendation framework. GenCDR first employs a \textbf{Domain-adaptive Tokenization} module, which generates disentangled semantic IDs for items by dynamically routing between a universal encoder and domain-specific adapters. Symmetrically, a \textbf{Cross-domain Autoregressive Recommendation} module models user preferences by fusing universal and domain-specific interests. Finally, a \textbf{Domain-aware Prefix-tree} enables efficient and accurate generation. Extensive experiments on multiple real-world datasets demonstrate that GenCDR significantly outperforms state-of-the-art baselines. Our code is available at \url{https://github.com/hupeiyu21/GenCDR}.

\end{abstract}

\section{Introduction}

Recommender systems have become indispensable tools for navigating the vast amount of information in modern online services, including e-commerce, social media, and content streaming~\cite{chen2024post, liu2024fine, li2025multi, guo2025higarment}. In real-world scenarios, users often interact across multiple, heterogeneous domains, creating rich behavioral data. Effectively leveraging these interactions for Cross-Domain Recommendation (CDR) has thus emerged as a critical challenge for improving recommendation accuracy and generalization~\cite{cross-domain-survey1, xiang2025harnessing}. However, the majority of existing CDR methods heavily rely on shared user or item identifiers (IDs) as the bridge for knowledge transfer~\cite{C2DSR,TriCDR}. This assumption often does not hold in practice, as many cross-domain scenarios, such as online content platforms and offline services, lack a strict alignment of user or item IDs, creating a significant bottleneck for traditional ID-based approaches~\cite{LLM4CDSR}.

\begin{figure}[t]
    \centering
    \includegraphics[width=0.99\linewidth]{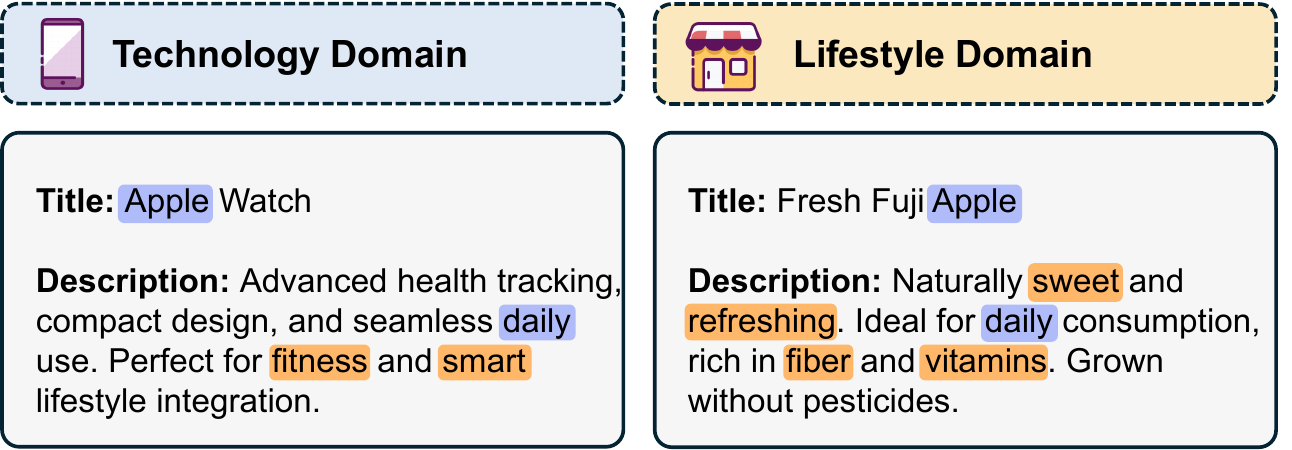}
    \caption{An ``Apple'' across Lifestyle vs. Technology domains. Blue: shared semantics (\textit{e.g., daily use}); Orange: domain-specific attributes (\textit{e.g., sweet, vitamins} for fresh apple; \textit{fitness, smart} for Apple Watch).}
    \label{fig:intro_example}
\end{figure}

The recent advancements in Large Language Models (LLMs) have provided a promising new paradigm for CDR, owing to their powerful capabilities in semantic understanding and sequence generation~\cite{llm-survey1, zeng2025numina}. Current research in this area can be broadly categorized into two main streams. The first stream utilizes LLMs as powerful feature enhancers, leveraging their world knowledge to generate richer representations for users and items, which are then fed into traditional recommendation models~\cite{llm-emb1, zhang2025llm}. The second stream treats the LLM as the core recommender, reformulating the CDR task as a natural language problem solved via prompting or parameter-efficient fine-tuning~\cite{LLM4CDSR}.

Despite this progress, current LLM-based CDR frameworks still face two fundamental challenges. The first is an \textbf{Item Tokenization Gap}, as traditional item indexing methods are ill-suited for LLMs in multi-domain scenarios~\cite{llm-survey1, li2023making, li2024survey}. The second is a \textbf{Domain Personalization Gap}, where existing methods struggle to effectively disentangle and model the dynamic interplay between universal interests and their domain-specific expressions~\cite{LLM4CDSR, zhang_adalora_2023, mo2025one}. For instance, as illustrated in Figure~\ref{fig:intro_example}, an ``Apple Watch'' in the technology domain and a fresh ``Apple'' in the lifestyle domain share a common semantic concept (Apple), but also possess highly distinct, domain-specific attributes (e.g., `health`, `fitness` vs. `sweet`, `vitamins`).

To address these challenges, we propose \textbf{GenCDR}, a novel \textbf{Generative Cross-Domain Recommendation framework based on Large Language Models}. Our work is motivated by a key insight: raw semantic information (e.g., text descriptions) is inherently transferable across domains, whereas item IDs are not. Inspired by the success of generative models in single-domain recommendation, we introduce the concept of discrete \textit{semantic IDs (SIDs)} to GenCDR, directly tackling the item tokenization dilemma (Challenge a). Furthermore, to address the lack of domain personalization (Challenge b), GenCDR features two core modules: a \textbf{Domain-adaptive Tokenization} module and a symmetric \textbf{Cross-Domain Autoregressive Recommendation} module. These are designed to systematically disentangle and dynamically fuse universal and domain-specific knowledge at both the item and user levels, respectively.

The main contributions of this paper are summarized as follows:
\begin{itemize}
    \item We propose a novel generative cross-domain recommendation framework, GenCDR. To the best of our knowledge, this is the first work to introduce the generative semantic ID paradigm into LLM-based cross-domain recommendation, effectively resolving the long-standing item tokenization dilemma.
    \item We systematically design a \textbf{Domain-adaptive Tokenization} module that dynamically disentangles and precisely models the universal and item-wise domain-specific knowledge at the semantic level.
    \item We design a symmetric and collaborative \textbf{Cross-Domain Autoregressive Recommendation} module that effectively disentangles and fuses universal and user-wise domain-specific interests during the recommendation process.
    \item We propose a \textbf{Domain-aware Prefix-tree} based decoding strategy to ensure efficient and accurate generation in cross-domain scenarios.
    \item Extensive experiments on multiple real-world cross-domain datasets demonstrate that GenCDR significantly outperforms existing state-of-the-art methods in terms of both accuracy and generalization.
\end{itemize}

\section{Related Work}

\textbf{Cross-Domain Sequential Recommendation} This task seeks to model a user's evolving interests across multiple domains by transferring knowledge from their diverse interaction sequences~\cite{overview}.  Mainstream approaches often rely on collaborative item IDs, using architectures like gating mechanisms, attention modules~\cite{SASRec, lu2025dmmd4sr, cui2025multi}, or Graph Neural Networks (GNNs)~\cite{liu2024unihr, li2024mhhcr, C2DSR} to fuse and transfer knowledge, frequently enhancing the representations with contrastive learning objectives~\cite{TriCDR, contrastive-learning}.  Recognizing the limitation of purely ID-based signals, a more recent trend has started to incorporate richer semantic information by leveraging features from pre-trained language models~\cite{LLM4CDSR, tokenCDSR}.  However, how to effectively integrate these semantics into a unified generative framework, while explicitly disentangling shared and domain-specific knowledge, remains a significant and open challenge.

\textbf{Generative Recommendation}. The paradigm of generative recommendation recasts the task from ranking to an autoregressive sequence generation problem, where Transformer-based models predict sequences of semantic item IDs~\cite{petrov2023generative, hou2025actionpiece}. The construction of these IDs is a critical research area, with key approaches including content-based tokenization via vector quantization~\cite{li2025semantic} (e.g., RQ-VAE~\cite{tiger}), structure-aware methods using hierarchical clustering~\cite{si2024generative}, and embedding collaborative signals directly into the tokenization process~\cite{mo2024min}. However, these techniques have been developed almost exclusively for single-domain datasets~\cite{zheng2025universal}, leaving their application to complex, multi-domain environments as an open research question.

\textbf{Large Language Models for Recommendation} Large Language Models (LLMs) are integrated into recommender systems in two main ways: either as auxiliary components that enhance traditional models by providing rich semantic features or data augmentation~\cite{sun_large_2024, yin2025unleash, zhang2025mitigating, yuan2025towards}, or as core generative engines that reformulate recommendation as a task of autoregressively predicting item IDs~\cite{tiger, zheng_lc-rec_2024, lin_bridging_2024-1}. Fine-tuning on recommendation datasets, often with parameter-efficient techniques (PEFT) like LoRA~\cite{hu2022lora}, is a crucial step to align these models for recommendation tasks~\cite{bao_tallrec_2023, liu_enhancing_2025, zhang_adalora_2023}. However, existing work has predominantly focused on single-domain applications, leaving the challenge of effective knowledge transfer and representation across heterogeneous domains largely unaddressed.

\section{Problem Formulation}
\label{sec:problem_formulation}

Let $\mathcal{U}$ be the set of users, $\mathcal{D}$ be the set of domains, and $\mathcal{I}_d$ be the item set for each domain $d \in \mathcal{D}$. The total item set is $\mathcal{I} = \bigcup_{d \in \mathcal{D}} \mathcal{I}_d$. For each user $u \in \mathcal{U}$, their historical interactions in a domain $d$ are represented as a chronological sequence $S_d^u = (i_1, \dots, i_t)$, where $i_k \in \mathcal{I}_d$. The user's complete cross-domain historical profile, $\mathcal{H}^u$, is the collection of all their single-domain sequences: $\mathcal{H}^u = \{ S_d^u \mid d \in \mathcal{D}_u \}$, where $\mathcal{D}_u \subseteq \mathcal{D}$ is the set of domains user $u$ has interacted with.

The task of Cross-Domain Sequential Recommendation (CDSR) is to predict the next item $i_{\text{target}}^u$ that a user $u$ is most likely to interact with in a target domain $d_t \in \mathcal{D}$, given their entire historical profile $\mathcal{H}^u$. The objective is to learn a generative model parameterized by $\theta$ that maximizes the log-likelihood of the held-out target items:
\begin{equation}
\mathcal{L} = \sum_{u \in \mathcal{U}} \log P(i_{\text{target}}^u | \mathcal{H}^u; \theta).
\end{equation}

\section{Methodology}

\begin{figure*}[t]
  \centering
  \includegraphics[width=0.99\textwidth]{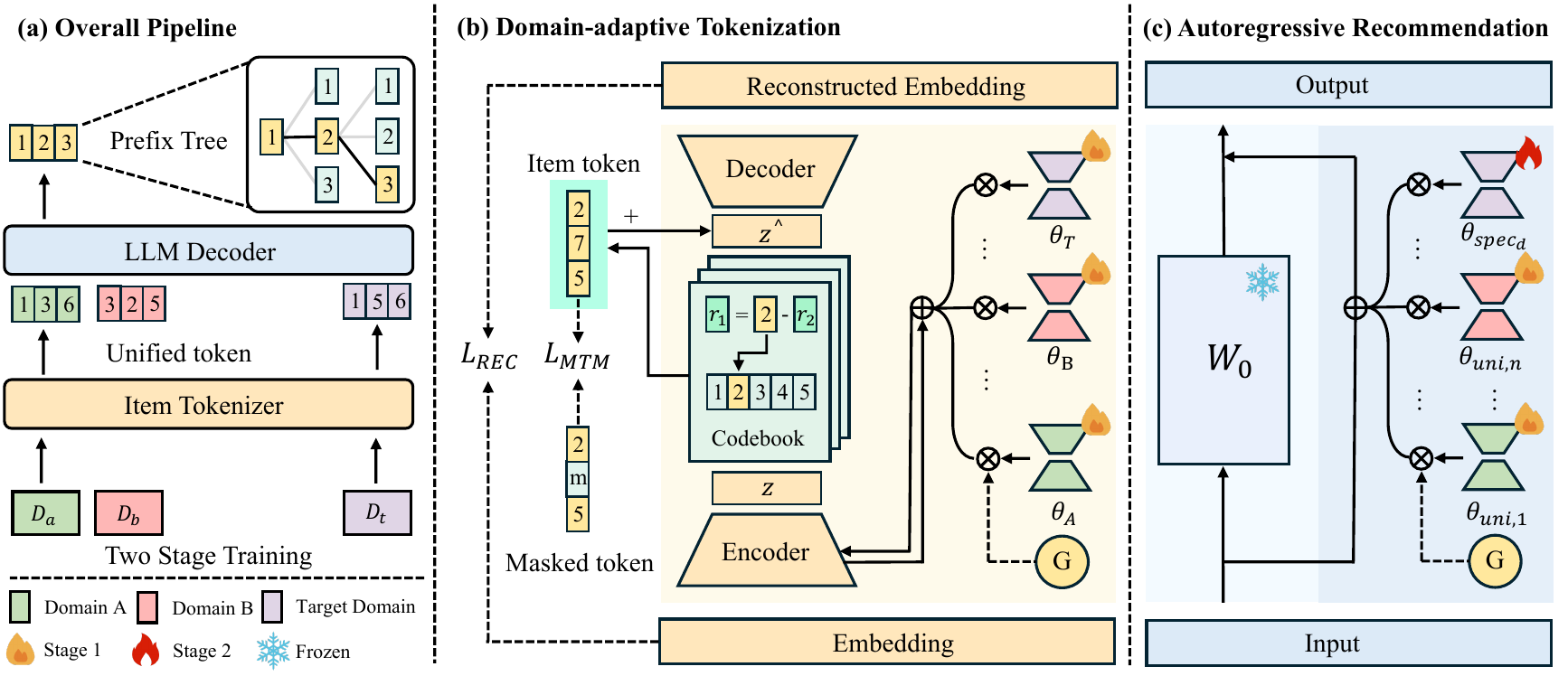}
  \caption{The architecture of our GenCDR framework. (a) The two-stage pipeline comprising the tokenization and recommendation modules. (b) The detailed structure of the Domain-adaptive Tokenization module, featuring a hierarchical adapter system with dynamic routing. (c) The symmetric architecture of the Cross-Domain Autoregressive Recommendation module.}
  \label{fig:architecture}
\end{figure*}

To address the key challenges of item tokenization and domain personalization in cross-domain recommendation, we propose \textbf{GenCDR}, a novel generative framework. As illustrated in Figure~\ref{fig:architecture}, our framework is composed of three core modules: a \textbf{Domain-adaptive Tokenization} module, a \textbf{Cross-Domain Autoregressive Recommendation} module, and a \textbf{Domain-aware Prefix-tree} for efficient inference.

\subsection{Domain-adaptive Tokenization}
\label{ssec:tokenization_module}

To empower large language models (LLMs) with the ability to process items from diverse domains, we introduce a \textit{Domain-adaptive Tokenization} module. This module generates unified SIDs for items, balancing domain-agnostic universal semantics with domain-specific discriminative features to yield expressive representations for generative recommendation tasks. The SIDs are designed to exhibit three critical properties: (i) \textit{Semantic Richness}: capturing comprehensive item semantics; and (ii) \textit{Semantic Similarity}: ensuring similar items across domains share comparable IDs.

\subsubsection{Domain-Universal Semantic Token Generation}
\label{sssec:universal_encoder}
To establish a unified semantic foundation for knowledge transfer, we propose a Universal Discrete Semantic Encoder based on a Residual-Quantized Variational Autoencoder (RQ-VAE) framework~\cite{rqvae}. The RQ-VAE, consisting of an encoder $E$, a decoder $D$, and $M$ codebooks, is pre-trained on the textual features of all items. It converts an item's feature embedding $\mathbf{x}$ into a sequence of discrete codes $\mathbf{c} = (c_0, \dots, c_{M-1})$.
The model encodes $\mathbf{x}$ to a latent representation $\mathbf{z} = E(\mathbf{x})$, with initial residual $\mathbf{r}_0 = \mathbf{z}$. For each level $d = 0$ to $M-1$, $\mathbf{r}_d$ is quantized to the nearest codebook vector $\mathbf{e}_{c_d}$ from codebook $\mathcal{C}_d$, where $c_d = \arg\min_k ||\mathbf{r}_d - \mathbf{e}_k||^2$, and the next residual is $\mathbf{r}_{d+1} = \mathbf{r}_d - \mathbf{e}_{c_d}$. The quantized $\hat{\mathbf{z}} := \sum_{d=0}^{M-1} \mathbf{e}_{c_d}$ is decoded to $\hat{\mathbf{x}} = D(\hat{\mathbf{z}})$.

The model is optimized via a joint objective function. This objective includes a standard reconstruction loss $\mathcal{L}_{\text{REC}} = || \mathbf{x} - \hat{\mathbf{x}} ||^2$ and a quantization loss $\mathcal{L}_{Q}$ that aligns the encoder's output with the codebook vectors using commitment terms~\cite{quantiloss}:
\begin{equation}
    \mathcal{L}_{Q} = \sum_{d=0}^{M-1} ||\text{sg}(\mathbf{r}_d) - \mathbf{e}_{c_d}|^2 + \beta ||\mathbf{r}_d - \text{sg}(\mathbf{e}_{c_d})||^2.
\end{equation}
To further ensure the codes are contextually coherent, we add a Masked Token Modeling (MTM) loss, which trains the model to predict masked codes $c_i$ from their surrounding context $S_{\text{masked}}$:
\begin{equation}
\mathcal{L}_{\text{MTM}} = - \mathbb{E}_{\mathbf{x} \sim \mathcal{X}, I_{\text{mask}}} \left[ \sum_{i \in I_{\text{mask}}} \log P(c_i \mid S_{\text{masked}}; \theta_{\text{ctx}}) \right].
\end{equation}
The total pre-training loss, $\mathcal{L}_{\text{pretrain}} = \mathcal{L}_{\text{REC}} + \mu \mathcal{L}_{Q} + \lambda \mathcal{L}_{\text{MTM}}$, guides the model to learn universal semantic tokens that are both representative and contextually aware. Upon completion, the universal encoder and codebooks are frozen.

\subsubsection{Domain-specific Semantic Token Adapters}
\label{sssec:domain_adapter}

While the universal encoder establishes a domain-agnostic semantic foundation, it may not fully capture domain-specific discriminative feature, such as visual aesthetics in videos or narrative styles in books. To address this, we introduce domain-specific semantic adapters that refine universal representations in a parameter-efficient manner, enhancing their relevance for each domain.

We leverage Low-Rank Adaptation (LoRA)~\cite{hu2022lora} to achieve this. For each domain $d \in \mathcal{D}$, a lightweight LoRA module is introduced, comprising low-rank matrices $B_d \in \mathbb{R}^{d_{\text{out}} \times r}$ and $A_d \in \mathbb{R}^{r \times d_{\text{in}}}$, where $r \ll \min(d_{\text{in}}, d_{\text{out}})$. These matrices augment the frozen weights $W_0 \in \mathbb{R}^{d_{\text{out}} \times d_{\text{in}}}$ of the universal encoder $E$, modifying the forward pass as:
\begin{equation}
h_{\text{out}} = W_0 h_{\text{in}} + B_d A_d h_{\text{in}}.
\end{equation}

Denoting the adapted encoder as $E_{\theta_d}$ with trainable parameters $\theta_d = \{B_d, A_d\}$ for domain $d$, we fine-tune $\theta_d$ in a second training phase. For each item embedding $\mathbf{x}$ from domain $d$, we minimize a self-supervised reconstruction loss:
\begin{equation}
\mathcal{L}_{\text{adapter}} = \mathbb{E}_{\mathbf{x} \sim \mathcal{X}_d} \left[ \|\mathbf{x} - D(Q(E_{\theta_d}(\mathbf{x})))\|_2^2 \right],
\end{equation}
where $Q$ and decoder $D$ remain frozen. This approach ensures domain-specific refinements with minimal additional parameters, enabling efficient adaptation to diverse domains.

\subsubsection{Item-level Dynamic Semantic Routing Network}
\label{sssec:routing_network}

To effectively integrate universal and domain-specific representations, we propose an Item-level Dynamic Semantic Routing Network that adaptively balances these representations on a per-item basis. This approach mitigates the risk of negative transfer inherent in static fusion strategies by dynamically determining the contribution of general cross-domain and domain-specific semantics for each item.

The routing network, denoted $R_\phi$ with parameters $\phi$, is a lightweight neural network (e.g., a multi-layer perceptron) that takes an item's embedding $\mathbf{x}$ as input and produces a gating weight $\alpha \in [0, 1]$. For an item from domain $d$, we compute two latent representations prior to quantization: the universal representation $\mathbf{z}_{\text{uni}} = E(\mathbf{x})$ from the frozen universal encoder, and the domain-specific representation $\mathbf{z}_{\text{spec}} = E_{\theta_d}(\mathbf{x})$ from the adapted encoder. The router calculates:
\begin{align}
\alpha &= \sigma(R_\phi(\mathbf{x})), \\
\mathbf{z}_{\text{fused}} &= (1 - \alpha) \cdot \mathbf{z}_{\text{uni}} + \alpha \cdot \mathbf{z}_{\text{spec}},
\end{align}
where $\sigma(\cdot)$ is the sigmoid function. The fused representation $\mathbf{z}_{\text{fused}}$ is then quantized and decoded.

To promote disentangled representations and prevent overfitting, we regularize the router using the Variational Information Bottleneck (VIB) principle~\cite{alemi2016deep}. The VIB loss minimizes the information the router extracts from $\mathbf{x}$, ensuring only essential features influence the routing decision. This is enforced via a KL-divergence term:
\begin{equation}
\mathcal{L}_{\text{VIB}} = D_{\text{KL}}(q(\mathbf{z}_r \mid \mathbf{x}) \parallel p(\mathbf{z}_r)),
\end{equation}
where $q(\mathbf{z}_r \mid \mathbf{x})$ is the router’s internal representation distribution, and $p(\mathbf{z}_r)$ is a prior (e.g., standard normal). This loss is incorporated into the second-phase training objective, enabling a balanced fusion of shared and domain-specific knowledge.

\subsection{Cross-Domain Autoregressive Recommendation}
\label{ssec:autoregressive_module}

Leveraging the unified SIDs produced by the Domain-adaptive Tokenization module, this component models intricate temporal patterns in user interaction sequences to enable personalized cross-domain recommendations. We introduce a parameter-efficient, two-phase fine-tuning strategy that acknowledges the multifaceted nature of user interests (e.g., brand preferences or category affinities). In the initial phase, a mixture of diverse LoRA adapters is trained on aggregated data from all domains to capture transferable, domain-agnostic interest patterns. The subsequent phase fine-tunes domain-specific LoRA adapters, with a dynamic routing network facilitating adaptive fusion of universal and specialized knowledge during inference.

\subsubsection{Universal Interest Modeling Network}
\label{sssec:universal_interest}

To model the diverse facets of user interests across domains, we develop a Universal Interest Modeling Network. This is achieved by enhancing a pre-trained large language model (LLM) with a mixture of multiple Low-Rank Adaptation (LoRA) adapters~\cite{li2024mixlora, zhang2024goal}. This collection of adapters is trained jointly to capture distinct, transferable behavioral patterns. The parameters of the $i$-th universal expert are denoted as $\theta_{\text{uni}, i}$. The complete set of these parameters, $\Theta_{\text{uni}} = \{\theta_{\text{uni}, 1}, \dots, \theta_{\text{uni}, N}\}$, represents all trainable weights for the universal module $G$.

The input to this network consists of sequences of cross-domain SIDs, $S^u = (c_1^u, c_2^u, \dots, c_t^u)$. In the initial fine-tuning phase, we optimize the universal parameters $\Theta_{\text{uni}}$ using a standard autoregressive objective, predicting the next semantic ID given the preceding sequence. The training loss is defined as:
\begin{equation}
\mathcal{L}_{\text{uni}} = - \sum_{u \in \mathcal{U}} \sum_{k=1}^{|S^u|-1} \log P(c_{k+1}^u \mid c_{\leq k}^u; \theta_{\text{LLM}}, \Theta_{\text{uni}}),
\label{eq:uni_loss}
\end{equation}
where $\theta_{\text{LLM}}$ denotes the frozen LLM parameters. After this phase, both $\theta_{\text{LLM}}$ and the universal parameter set $\Theta_{\text{uni}}$ are fixed, forming the Universal Interest Modeling Network that serves as the basis for domain-specific adaptation.

\subsubsection{Domain-specific Interest Adaptation}
\label{sssec:specific_interest}

While the Universal Interest Modeling Network captures general user preferences, domain-specific nuances require tailored modeling. To address this, we introduce a second fine-tuning phase to train domain-specific LoRA adapters, enabling the model to adapt to the unique characteristics of each domain.

For each domain $d \in \mathcal{D}$, we augment the frozen model with a dedicated, trainable LoRA adapter, denoted $\theta_{\text{spec}_d}$. During this phase, both the base LLM parameters $\theta_{\text{LLM}}$ and the universal parameters $\Theta_{\text{uni}}$ remain fixed. The training focuses solely on domain-specific interaction sequences $S_d^u$ from users $u \in \mathcal{U}_d$. We optimize $\theta_{\text{spec}_d}$ by minimizing the autoregressive loss:
\begin{equation}
\label{eq:spec_loss}
\mathcal{L}_{\text{spec}_d} = -\sum_{u \in \mathcal{U}_d} \sum_{k=1}^{|S_d^u|-1}  
\log P(c_{k+1}^u \mid c_{\leq k}^u; \theta_{\text{LLM}}, \Theta_{\text{uni}}, \theta_{\text{spec}_d}).
\end{equation}
This approach enables the model to efficiently learn domain-specific interest patterns, setting the stage for dynamic integration during inference.

\subsubsection{User-level Dynamic Interest Routing Network}
\label{sssec:user_routing}

Symmetrically to the item-level router, we employ a VIB-regularized User-level Dynamic Interest Routing Network to prevent negative transfer during inference. This lightweight gate takes the user's history representation $\mathbf{h}_t$ as input to compute a dynamic weight $\gamma \in [0, 1]$. This weight fuses the probability distributions from the universal model ($P_{\text{uni}}$) and the domain-adapted model ($P_{\text{spec}}$) as follows:
\begin{equation}
P_{\text{final}}(i \mid S^u) = (1 - \gamma) \cdot P_{\text{uni}}(i \mid S^u) + \gamma \cdot P_{\text{spec}}(i \mid S^u).
\label{eq:fusion_pred}
\end{equation}
Here, $P_{\text{uni}}$ is the output distribution from the frozen universal network (parameterized by $\Theta_{\text{uni}}$), while $P_{\text{spec}}$ is from the network augmented with domain-specific adapters (parameterized by $\Theta_{\text{uni}}$ and $\theta_{\text{spec}_d}$). The VIB regularization on the router ensures the fusion logic is efficient and robust.

\subsubsection{Inference – Domain-aware Prefix-tree}
\label{sssec:constrained_generation}

To ensure efficient and valid semantic ID generation, we propose a Domain-aware Prefix-tree mechanism that mitigates the limitations of standard autoregressive decoding, such as computational inefficiency and invalid ID outputs. For each domain $d \in \mathcal{D}$, we construct an offline prefix tree $T_d$ encoding all valid semantic ID sequences produced by the Domain-adaptive Tokenization module. During inference, given a target domain $d_t$, the corresponding tree $T_{d_t}$ guides the generation process. At each decoding step $k$, the tree identifies a valid subset of next codes $V_{\text{valid}}(s_{k-1}) \subset C_k$ based on the current prefix $s_{k-1}$. The LLM’s predictions are constrained to this subset using a masked softmax:
\begin{equation}
P(c_k \mid s_{k-1}, T_{d_t}) = \frac{\exp(z_k)}{\sum_{c' \in V_{\text{valid}}(s_{k-1})} \exp(z_{c'})},
\end{equation}
where $z_k$ are the LLM’s logits. This approach ensures valid sequence generation while significantly reducing computational overhead, enhancing the efficiency of the recommendation process.

\section{Experiments}

In this section, we conduct extensive experiments on several public datasets to evaluate our proposed model. The experiments are designed to answer the following key Research Questions (RQs):

\begin{itemize}
    \item \textbf{(Effectiveness) RQ1:} How does our proposed model perform against state-of-the-art single-domain and cross-domain recommendation baselines?

    \item \textbf{(Ablation) RQ2:} What is the contribution of each key component in our framework?

    \item \textbf{(Analysis) RQ3:} Can our framework learn visually separable representations for universal and domain-specific knowledge?

    \item \textbf{(Sensitivity) RQ4:} How does our model's performance change with respect to key hyper-parameter settings?

    \item \textbf{(Efficiency) RQ5:} How efficient is our GenCDR framework in terms of training cost and inference scalability?

\end{itemize}

\subsection{Experimental Setup}

\textbf{Datasets.} We experiment on three cross-domain dataset pairs, each reflecting a distinct real-world scenario: \textbf{Sports-Clothing} (Leisure), \textbf{Phones-Electronics} (Technology), and \textbf{Books-Movies} (Entertainment). The first two pairs are derived from the public Amazon product review dataset~\cite{amazon-dataset}, while the third is collected from Douban~\cite{douban1, douban2}. 

Following ~\cite{tiger, s3rec}, we treat users' historical reviews as interactions arranged chronologically. We use the leave-last-out evaluation protocol ~\cite{SASRec, leaveout}, where the last item is for testing, and the second-to-last for validation. Table~\ref{tab:dataset-stats} shows dataset statistics.

\begin{table}[t]
\centering
\setlength{\tabcolsep}{1pt}  

\small
\begin{tabularx}{\linewidth}{>{\centering\arraybackslash}l 
                                  >{\centering\arraybackslash}X 
                                  >{\centering\arraybackslash}X 
                                  >{\centering\arraybackslash}X 
                                  >{\centering\arraybackslash}X 
                                  >{\centering\arraybackslash}X}
\toprule
{\scriptsize\textbf{Dataset}} & {\scriptsize\textbf{\#Users}} & {\scriptsize\textbf{\#Items}} & {\scriptsize\textbf{\#Interactions}} & {\scriptsize\textbf{Sparsity}} & {\scriptsize\textbf{Overlap}} \\

\midrule
Sports   & 35,598  & 18,357  & 296,337 & 99.95\% & \multirow{2}{*}{\makecell{1.73\%\\(704)}} \\
Clothing & 39,387  & 23,033  & 278,677 & 99.97\% &                                             \\
\midrule
Phones       & 27,879  & 10,429  & 194,439 & 99.93\% & \multirow{2}{*}{\makecell{0.55\%\\(404)}} \\
Electronics  &192,403  & 63,001  & 1,689,188 & 99.99\% &                                             \\
\midrule
Books    & 1,713   & 8,601   & 104,295  & 99.29\% & \multirow{2}{*}{\makecell{7.48\%\\(2,058)}} \\
Movies   & 2,628   & 20,964  & 1,249,016  & 97.73\% &                                             \\
\bottomrule
\end{tabularx}
\caption{Statistics of the datasets used in our experiments. Item overlap and sparsity are computed after merging.}
\label{tab:dataset-stats}
\end{table}

\begin{table*}[!htbp]
\centering

\newcolumntype{C}{>{\centering\arraybackslash}X} 
\setlength{\tabcolsep}{2pt} 
\small
\renewcommand{\arraystretch}{0.9} 

\begin{tabularx}{\textwidth}{l|l|l|CCC|CCC|CCC|C}
\toprule
\multicolumn{2}{c|}{\textbf{Dataset}} & \textbf{Metric} 
  & \multicolumn{3}{c|}{\textbf{SDSR}} 
  & \multicolumn{3}{c|}{\textbf{GenRec}} 
  & \multicolumn{3}{c|}{\textbf{CDSR}} 
  & \textbf{Ours} \\
\cmidrule(lr){1-2} \cmidrule(lr){4-6} \cmidrule(lr){7-9} \cmidrule(lr){10-12} \cmidrule(lr){13-13}
\textbf{Scene} & \textbf{Domain} &  & 
{\fontsize{7pt}{8pt}\selectfont Bert4Rec} & 
{\fontsize{7pt}{8pt}\selectfont SASRec} & 
{\fontsize{7pt}{8pt}\selectfont STOSA} & 
{\fontsize{7pt}{8pt}\selectfont VQ-Rec} & 
{\fontsize{7pt}{8pt}\selectfont TIGER} & 
{\fontsize{7pt}{8pt}\selectfont HSTU} & 
{\fontsize{7pt}{8pt}\selectfont C2DSR} & 
{\fontsize{7pt}{8pt}\selectfont TriCDR} & 
{\fontsize{7pt}{8pt}\selectfont LLM4CDSR} & 
\textbf{{\fontsize{7pt}{8pt}\selectfont GenCDR}} \\
\midrule

\multirow{8}{*}{\rotatebox{90}{\textbf{Leisure}}} & \multirow{4}{*}{Sports}
& R@5  & 0.0188 & 0.0197 & 0.0236 & 0.0261 & \underline{0.0267} & 0.0254 & 0.0265 & 0.0266 & 0.0263 & \textbf{0.0274} \\
& & N@5  & 0.0121 & 0.0126 & 0.0162 & 0.0238 & 0.0244 & 0.0241 & 0.0253 & 0.0255 & \underline{0.0257} & \textbf{0.0261} \\
& & R@10 & 0.0325 & 0.0334 & 0.0346 & 0.0389 & \underline{0.0397} & 0.0381 & 0.0395 & 0.0396 & 0.0398 & \textbf{0.0403} \\
& & N@10 & 0.0169 & 0.0173 & \underline{0.0283} & 0.0281 & \textbf{0.0287} & 0.0277 & 0.0258 & 0.0259 & 0.0260 & 0.0262 \\
\cmidrule(l){2-13}
& \multirow{4}{*}{Clothing}
& R@5  & 0.0128 & 0.0132 & 0.0162 & 0.0171 & 0.0173 & 0.0175 & 0.0172 & 0.0174 & \underline{0.0176} & \textbf{0.0181} \\
& & N@5  & 0.0078 & 0.0081 & 0.0119 & 0.0129 & 0.0125 & 0.0132 & 0.0158 & 0.0161 & \underline{0.0163} & \textbf{0.0167} \\
& & R@10 & 0.0219 & 0.0227 & 0.0223 & 0.0248 & 0.0241 & 0.0253 & 0.0255 & 0.0258 & \underline{0.0261} & \textbf{0.0265} \\
& & N@10 & 0.0105 & 0.0108 & 0.0135 & 0.0170 & 0.0167 & 0.0174 & 0.0191 & 0.0194 & \underline{0.0196} & \textbf{0.0203} \\
\midrule

\multirow{8}{*}{\rotatebox{90}{\textbf{Technology}}} & \multirow{4}{*}{Phones} 
& R@5  & 0.0331 & 0.0345 & 0.0415 & 0.0411 & 0.0423 & 0.0415 & 0.0428 & 0.0434 & \underline{0.0431} & \textbf{0.0436} \\
& & N@5  & 0.0215 & 0.0224 & 0.0283 & 0.0308 & 0.0315 & 0.0327 & 0.0392 & 0.0396 & \underline{0.0401} & \textbf{0.0411} \\
& & R@10 & 0.0524 & 0.0537 & 0.0618 & 0.0607 & 0.0613 & \underline{0.0615} & 0.0589 & 0.0593 & 0.0614 & \textbf{0.0621} \\
& & N@10 & 0.0278 & 0.0287 & 0.0346 & 0.0399 & 0.0406 & 0.0425 & 0.0493 & 0.0505 & \underline{0.0506} & \textbf{0.0512} \\
\cmidrule(l){2-13}
& \multirow{4}{*}{Electronics}
& R@5  & 0.0179 & 0.0186 & 0.0213 & 0.0219 & 0.0228 & 0.0232 & 0.0235 & \underline{0.0238} & 0.0237 & \textbf{0.0241} \\
& & N@5  & 0.0118 & 0.0122 & 0.0148 & 0.0211 & 0.0214 & 0.0226 & 0.0229 & \underline{0.0231} & 0.0230 & \textbf{0.0235} \\
& & R@10 & 0.0276 & 0.0285 & 0.0315 & 0.0318 & 0.0322 & 0.0328 & 0.0336 & \underline{0.0339} & 0.0338 & \textbf{0.0342} \\
& & N@10 & 0.0149 & 0.0154 & 0.0172 & 0.0262 & 0.0269 & 0.0271 & 0.0278 & \underline{0.0280} & 0.0279 & \textbf{0.0283} \\
\midrule

\multirow{8}{*}{\rotatebox{90}{\textbf{Entertainment}}} & 
\multirow{4}{*}{Books}
& R@5  & 0.0089 & 0.0093 & 0.0142 & 0.0175 & 0.0172 & \underline{0.0181} & 0.0152 & 0.0155 & 0.0161 & \textbf{0.0192} \\
& & N@5  & 0.0071 & 0.0076 & 0.0117 & 0.0178 & 0.0177 & \underline{0.0180} & 0.0143 & 0.0148 & 0.0153 & \textbf{0.0187} \\
& & R@10 & 0.0176 & 0.0182 & 0.0219 & 0.0224 & 0.0221 & \underline{0.0230} & 0.0205 & 0.0211 & 0.0216 & \textbf{0.0237} \\
& & N@10 & 0.0158 & 0.0164 & 0.0165 & 0.0201 & 0.0198 & \underline{0.0206} & 0.0182 & 0.0185 & 0.0189 & \textbf{0.0212} \\
\cmidrule(l){2-13}
& \multirow{4}{*}{Movies}
& R@5  & 0.1503 & 0.1542 & 0.1562 & 0.1680 & 0.1652 & \underline{0.1682} & 0.1588 & 0.1601 & 0.1613 & \textbf{0.1713} \\
& & N@5  & 0.1015 & 0.1047 & 0.1063 & 0.1182 & 0.1156 & \underline{0.1189} & 0.1092 & 0.1105 & 0.1149 & \textbf{0.1215} \\
& & R@10 & 0.1798 & 0.1825 & 0.1753 & 0.1922 & 0.1893 & \underline{0.1931} & 0.1854 & 0.1865 & 0.1878 & \textbf{0.1971} \\
& & N@10 & 0.1211 & 0.1265 & 0.1223 & 0.1261 & 0.1255 & \underline{0.1268} & 0.1203 & 0.1217 & 0.1225 & \textbf{0.1275} \\

\bottomrule
\end{tabularx}
\caption{Overall performance comparison on all datasets. R@K and N@K denote Recall and NDCG at cutoff K. Best results are in \textbf{bold}, and the best baseline results are \underline{underlined}. The $t$-tests showed significant performance improvements ($p \le 0.05$).}
\label{tab:all_domains_data_segmented_centered}

\end{table*}

\textbf{Baselines.} To comprehensively evaluate the effectiveness of our proposed GenCDR framework, we compare it with three representative categories of state-of-the-art models: 
(1) \textbf{Single-domain Sequential Recommendation (SDSR)}, 
(2) \textbf{Generative Recommendation Systems (GRS)}, and 
(3) \textbf{Cross-domain Sequential Recommendation (CDSR)}. 

For single-domain models such as \textbf{SASRec}~\cite{SASRec}, \textbf{BERT4Rec}~\cite{sun2019bert4rec}, and \textbf{STOSA}~\cite{fan2022stosa}, we follow their standard single-domain setups to ensure fair comparison. 
For generative and cross-domain models including \textbf{VQ-Rec}~\cite{vqrec}, \textbf{TIGER}~\cite{tiger}, \textbf{HSTU}~\cite{hstu}, \textbf{C2DSR}~\cite{C2DSR}, \textbf{TriCDR}~\cite{TriCDR}, and \textbf{LLM4CDSR}~\cite{LLM4CDSR}, 
we adopt their official multi-domain configurations to fully exploit their cross-domain transfer capability. 
All baselines are re-implemented and tuned under a unified PyTorch framework for consistency.

\begin{itemize}
    \item \textbf{Single-domain Sequential Recommendation (SDSR).}  
    \begin{itemize}
        \item \textbf{SASRec}~\cite{SASRec} employs a unidirectional Transformer to model users’ sequential preferences through self-attention, allowing it to highlight the most relevant past interactions when predicting the next item.
        \item \textbf{BERT4Rec}~\cite{sun2019bert4rec} extends BERT to recommendation by using a masked item prediction objective, enabling bidirectional context learning that captures both past and future dependencies.
        \item \textbf{STOSA}~\cite{fan2022stosa} introduces stochastic self-attention for long sequences, enhancing efficiency while incorporating self-supervised objectives for more robust item representations.
    \end{itemize}

    \item \textbf{Generative Recommendation Systems (GRS).}  
    \begin{itemize}
        \item \textbf{VQ-Rec}~\cite{vqrec} combines VQ-VAE-based tokenization and Transformer sequence modeling, mapping item embeddings to discrete codes before predicting the next item in code space.
        \item \textbf{TIGER}~\cite{tiger} enhances generative retrieval by optimizing item tokenization with collaborative constraints, producing semantic IDs that capture both content and user–item interaction signals.
        \item \textbf{HSTU}~\cite{hstu} proposes a hierarchical tokenization framework that encodes items at multiple semantic levels (from coarse to fine-grained), improving both generation accuracy and efficiency.
    \end{itemize}

    \item \textbf{Cross-domain Sequential Recommendation (CDSR).}   
    \begin{itemize}
        \item \textbf{C2DSR}~\cite{C2DSR} constructs a unified user–item interaction graph across domains and employs a GNN-based propagation mechanism with adaptive gating to regulate inter-domain knowledge transfer.
        \item \textbf{TriCDR}~\cite{TriCDR} utilizes triplet-based contrastive learning to align user embeddings across domains by minimizing cross-domain intra-user distances and maximizing inter-user separability.
        \item \textbf{LLM4CDSR}~\cite{LLM4CDSR} reformulates CDR as a text generation task, converting user histories and item attributes into textual prompts for LLMs to model implicit cross-domain semantic correlations.
    \end{itemize}
\end{itemize}

\textbf{Evaluation Metrics.} Following the standard practice in sequential recommendation literature~\cite{SASRec, tiger}, we adopt Recall@$K$ and NDCG@$K$ as our evaluation metrics, with $K$ set to 5 and 10. For each model, the checkpoint that achieves the best Recall@10 on the validation set is selected for the final testing phase.

\textbf{Implementation Details.} 
Our framework is implemented in PyTorch with Hugging Face PEFT for LoRA-based fine-tuning. The training of GenCDR consists of two main stages. 
In the first stage, we train the \textit{Domain-adaptive Tokenization} module: the RQ-VAE is pre-trained on all item embeddings using AdamW (lr=$1\times10^{-4}$, batch=512) for 100 epochs, followed by domain-specific LoRA adapters (rank=64, $\alpha$=32, dropout=0.05) fine-tuned for 50 epochs with lr=$5\times10^{-5}$. The router network is a two-layer MLP with 128 hidden units, trained jointly with a VIB regularization weight of $10^{-3}$. In the second stage, we fine-tune the \textit{Cross-Domain Autoregressive Recommendation} module using the Qwen2.5–7B backbone. We first train $N=4$ universal LoRA experts (rank=64, $\alpha$=128) on combined cross-domain data for 10 epochs, and then fine-tune domain-specific adapters for 10–20 epochs per domain. 
All models are optimized with AdamW (lr=$5\times10^{-5}$, batch=8) under mixed-precision (FP16) on NVIDIA H200 GPUs.

\subsection{Overall Performance (RQ1)}

The overall performance comparison of our proposed GenCDR against all baseline models is summarized in Table~\ref{tab:all_domains_data_segmented_centered}. The results show that our proposed GenCDR consistently and significantly outperforms all baseline models, demonstrating its overall superiority in the cross-domain sequential recommendation task.

An analysis of the baselines provides clear insights into this improvement. We observe that cross-domain (CDSR) models generally yield better results than traditional single-domain (SDSR) models, which validates the fundamental premise of leveraging cross-domain information. Furthermore, while generative (GenRec) models also show an advantage over SDSR baselines, their performance typically falls short of specialized CDSR models. This hierarchy of performance precisely motivates our work, as it suggests that simply applying existing generative models to cross-domain scenarios is a suboptimal strategy. Our GenCDR framework is specifically designed to bridge this gap by deeply integrating the generative paradigm with the unique challenges of cross-domain knowledge transfer, thereby achieving state-of-the-art performance.

\subsection{Ablation Study (RQ2)} \label{sec:rq2}

\begin{table*}[ht]
\centering
\small

\setlength{\tabcolsep}{8pt} 

\begin{tabular}{llcccc}
\toprule
\textbf{Category} & \textbf{Variant} & \textbf{Phones} & \textbf{Electronics} & \textbf{Sports} & \textbf{Clothing} \\
\midrule

\textbf{Full Model} & \textbf{GenCDR} & \textbf{0.0512} & \textbf{0.0283} & \textbf{0.0262} & \textbf{0.0203} \\
\midrule

\multirow{2}{*}{Tokenization} 
& w/o MTM & 0.0483~{\scriptsize ($\downarrow$5.7\%)} & 0.0267~{\scriptsize ($\downarrow$5.7\%)} & 0.0245~{\scriptsize ($\downarrow$6.5\%)} & 0.0190~{\scriptsize ($\downarrow$6.4\%)} \\
& w/o Adapter & 0.0466~{\scriptsize ($\downarrow$9.0\%)} & 0.0255~{\scriptsize ($\downarrow$9.9\%)} & 0.0238~{\scriptsize ($\downarrow$9.2\%)} & 0.0183~{\scriptsize ($\downarrow$9.9\%)} \\
\midrule

\multirow{3}{*}{Autoregressive Recommendation}
& w/o Specific Expert & 0.0448~{\scriptsize ($\downarrow$12.5\%)} & 0.0245~{\scriptsize ($\downarrow$13.4\%)} & 0.0226~{\scriptsize ($\downarrow$13.7\%)} & 0.0173~{\scriptsize ($\downarrow$14.8\%)} \\
& w/o Universal Experts & 0.0425~{\scriptsize ($\downarrow$17.0\%)} & 0.0232~{\scriptsize ($\downarrow$18.0\%)} & 0.0212~{\scriptsize ($\downarrow$19.1\%)} & 0.0162~{\scriptsize ($\downarrow$20.2\%)} \\
& w/o MoE Gate (Avg.) & 0.0475~{\scriptsize ($\downarrow$7.2\%)} & 0.0262~{\scriptsize ($\downarrow$7.4\%)} & 0.0242~{\scriptsize ($\downarrow$7.6\%)} & 0.0186~{\scriptsize ($\downarrow$8.4\%)} \\
\midrule

\multirow{1}{*}{Inference Strategy}
& w/o Prefix Tree & 0.0498~{\scriptsize ($\downarrow$2.7\%)} & 0.0274~{\scriptsize ($\downarrow$3.2\%)} & 0.0255~{\scriptsize ($\downarrow$2.7\%)} & 0.0198~{\scriptsize ($\downarrow$2.5\%)} \\

\bottomrule
\end{tabular}

\caption{Ablation study on GenCDR components across four datasets (NDCG@10). Values in parentheses denote the drop in performance compared to the full model.}
\label{tab:ablation}

\end{table*}

To dissect the contribution of each key design choice in our GenCDR framework, we conducted a thorough ablation study. The results, summarized in Table~\ref{tab:ablation}, unequivocally demonstrate that each component plays an integral role.

\begin{itemize}
    \item \textbf{Impact of Contextual Code Modeling.} 
    Removing the MTM loss (\textbf{w/o MTM}) degrades performance, confirming that learning the contextual "grammar" of the semantic codes is crucial, beyond simple reconstruction.

    \item \textbf{Impact of Item-specific Adaptation.} 
    Removing the item-specific adapter (\textbf{w/o Item Adapter}) degrades performance, validating the need for domain-specific item semantics.

    \item \textbf{Impact of the Specific Expert.} 
    Removing the domain-specific expert (\textbf{w/o Specific Expert}) significantly hurts performance, proving its crucial role in capturing fine-grained user preferences.

    \item \textbf{Impact of the Universal Experts.} 
    Removing all $N$ universal experts (\textbf{w/o Universal Experts}) causes a sharp performance drop, confirming that a shared cross-domain knowledge foundation is indispensable.

    \item \textbf{Impact of the MoE Gate.}
    Replacing the trainable MoE gate with simple averaging (\textbf{w/o MoE Gate}) hurts performance, highlighting the importance of a dynamic, context-aware selection of experts over naive fusion.

    \item \textbf{Impact of Constrained Decoding.} 
    Removing the prefix-tree constraint (\textbf{w/o Prefix Tree}) leads to a consistent performance drop, as it guarantees the generation of valid item IDs and prevents "hallucinated" recommendations.
\end{itemize}

\subsection{In-depth Analysis (RQ3)}

\begin{figure}[h]
    \centering
    \includegraphics[width=1\linewidth]{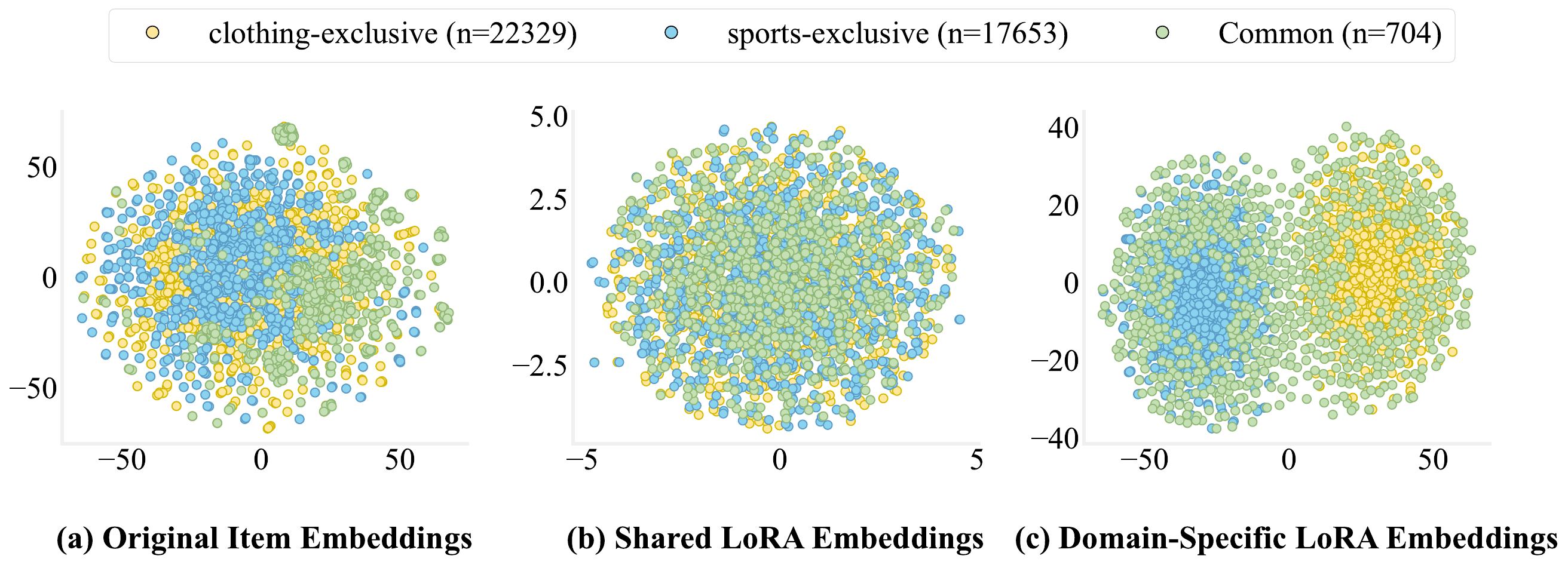}
    \caption{t-SNE visualization of item embeddings in three different settings. }
    \label{fig:tsne_visualization}
\end{figure}

To qualitatively assess our framework, we visualize the final item representations ($\mathbf{z}_{\text{fused}}$) using t-SNE in Figure~\ref{fig:tsne_visualization}. In (b), using only universal adapters, item embeddings from different domains are mixed together. In contrast, (c) shows our full GenCDR model with domain-specific adapters, where embeddings form clearly separated domain-specific clusters. This confirms the importance of domain-specific adaptation for learning disentangled representations.

\begin{figure}[h]
    \centering
    \includegraphics[width=1\linewidth]{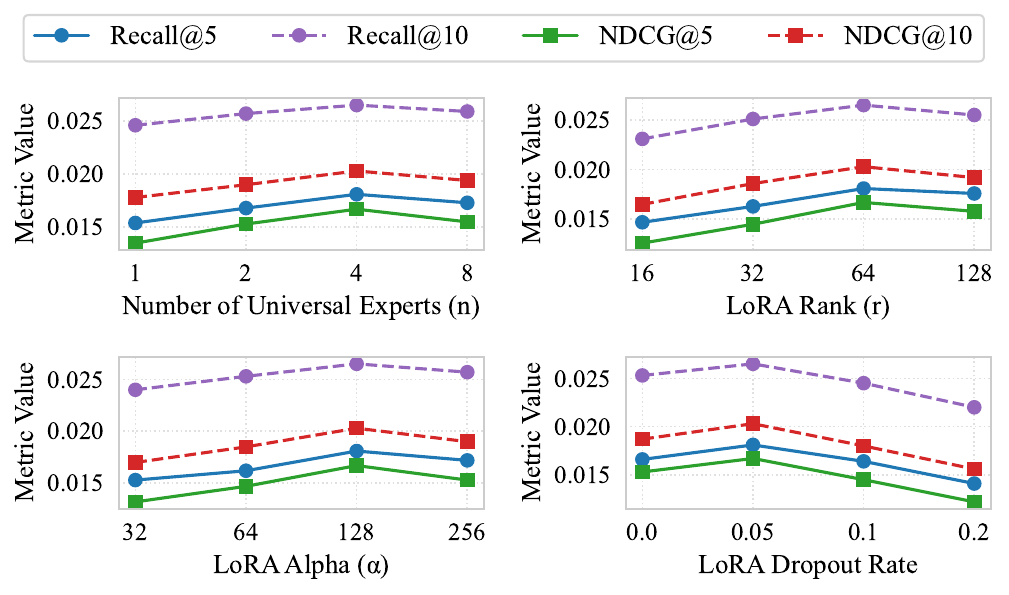}
    \caption{Sensitivity of LoRA fine-tuning to key hyper-parameters on the Cloth dataset.}
    \label{fig:hyper_sensitivity}
\end{figure}

\subsection{Hyper-parameter Analysis (RQ4)}

\begin{figure}[t]
    \centering
    \includegraphics[width=1\linewidth]{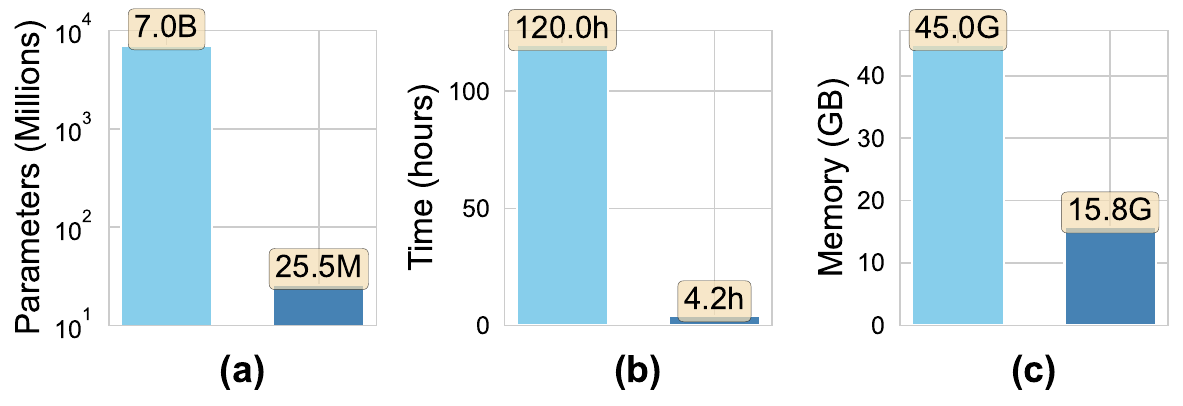}
    \caption{Comparison of training efficiency using the Qwen2.5-7B model. The plots show (a) trainable parameters (log scale), (b) training time, and (c) peak GPU memory for our LoRA-based GenCDR versus a Full Fine-Tuning (Full FT) version.}
    \label{fig:training_efficiency}
\end{figure}

We analyze the sensitivity of key hyperparameters on the \textbf{Cloth} dataset in Figure~\ref{fig:hyper_sensitivity}. The results for \textbf{Universal Experts ($N$)}, \textbf{LoRA Rank ($r$)}, and \textbf{Alpha ($\alpha$)} reveal clear optima (e.g., $N=4$, $r=64$), beyond which performance declines due to overfitting. A small \textbf{LoRA Dropout Rate} (0.05) offers effective regularization. These findings highlight a balanced trade-off between capacity and generalization, demonstrating the framework’s robustness and tunability.

\subsection{Analysis of Efficiency (RQ5)}
\label{ssec:efficiency_analysis}

\textit{Training Efficiency.} As shown in Figure~\ref{fig:training_efficiency}, our LoRA-based fine-tuning greatly reduces trainable parameters, training time, and GPU memory compared to full fine-tuning.

\begin{figure}[t]
    \centering
    \includegraphics[width=0.9\linewidth]{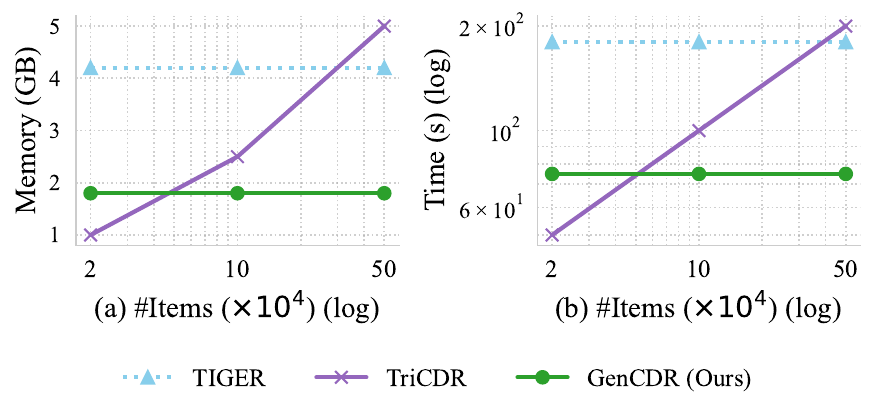}
    \caption{Comparison of runtime memory and inference time w.r.t. the item pool size for TriCDR, TIGER, and GenCDR (Qwen2.5-0.5B).}
    \label{fig:inference_efficiency}
\end{figure}

\textit{Inference Efficiency and Scalability.} Furthermore, Figure~\ref{fig:inference_efficiency} demonstrates GenCDR's superior scalability, as its inference cost remains constant regardless of the item pool size due to our prefix-tree constrained generative architecture.

\section{Conclusion}

In this paper, we addressed the critical challenges of item tokenization and domain personalization in LLM-based cross-domain recommendation. We proposed \textbf{GenCDR}, a novel generative framework that systematically tackles these issues. Our approach introduces a \textbf{Domain-adaptive Tokenization} module to create hybrid SIDs and a symmetric \textbf{Cross-Domain Autoregressive Recommendation} module to dynamically model user interests by fusing universal and specific knowledge. Furthermore, a prefix-tree mechanism ensures efficient and valid inference. Experiments show that GenCDR achieves superior effectiveness and scalability, and future work will explore incorporating multimodal features for richer representations.

\section{Acknowledgements}
This research was supported in part by the National Natural Science Foundation of China (Grant No. 72401232), the Natural Science Foundation of the Jiangsu Higher Educational Institution of China (Grant No. 23KJB520037), and the XJTLU Research Development Fund under RDF-21-01-053.

\bibliography{aaai2026}

@article{tiger,
  title={Recommender systems with generative retrieval},
  author={Rajput, Shashank and Mehta, Nikhil and Singh, Anima and Hulikal Keshavan, Raghunandan and Vu, Trung and Heldt, Lukasz and Hong, Lichan and Tay, Yi and Tran, Vinh and Samost, Jonah and others},
  journal={Advances in Neural Information Processing Systems},
  volume={36},
  pages={10299--10315},
  year={2023}
}

@inproceedings{si2024generative,
  title={Generative retrieval with semantic tree-structured identifiers and contrastive learning},
  author={Si, Zihua and Sun, Zhongxiang and Chen, Jiale and Chen, Guozhang and Zang, Xiaoxue and Zheng, Kai and Song, Yang and Zhang, Xiao and Xu, Jun and Gai, Kun},
  booktitle={Proceedings of the 2024 Annual International ACM SIGIR Conference on Research and Development in Information Retrieval in the Asia Pacific Region},
  pages={154--163},
  year={2024}
}

@article{petrov2023generative,
  title={Generative sequential recommendation with gptrec},
  author={Petrov, Aleksandr V and Macdonald, Craig},
  journal={arXiv preprint arXiv:2306.11114},
  year={2023}
}

@article{hou2025actionpiece,
  title={ActionPiece: Contextually Tokenizing Action Sequences for Generative Recommendation},
  author={Hou, Yupeng and Ni, Jianmo and He, Zhankui and Sachdeva, Noveen and Kang, Wang-Cheng and Chi, Ed H and McAuley, Julian and Cheng, Derek Zhiyuan},
  journal={arXiv preprint arXiv:2502.13581},
  year={2025}
}

@inproceedings{li2025semantic,
  title={Semantic convergence: Harmonizing recommender systems via two-stage alignment and behavioral semantic tokenization},
  author={Li, Guanghan and Zhang, Xun and Zhang, Yufei and Yin, Yifan and Yin, Guojun and Lin, Wei},
  booktitle={Proceedings of the AAAI Conference on Artificial Intelligence},
  volume={39},
  number={11},
  pages={12040--12048},
  year={2025}
}

@article{zheng2025universal,
  title={Universal Item Tokenization for Transferable Generative Recommendation},
  author={Zheng, Bowen and Lu, Hongyu and Chen, Yu and Zhao, Wayne Xin and Wen, Ji-Rong},
  journal={arXiv preprint arXiv:2504.04405},
  year={2025}
}

@article{overview,
  title={A survey on cross-domain sequential recommendation},
  author={Chen, Shu and Xu, Zitao and Pan, Weike and Yang, Qiang and Ming, Zhong},
  journal={arXiv preprint arXiv:2401.04971},
  year={2024}
}

@inproceedings{SASRec,
  title={Self-attentive sequential recommendation},
  author={Kang, Wang-Cheng and McAuley, Julian},
  booktitle={2018 IEEE international conference on data mining (ICDM)},
  pages={197--206},
  year={2018},
  organization={IEEE}
}

@inproceedings{contrastive-learning,
  title={Contrastive learning for sequential recommendation},
  author={Xie, Xu and Sun, Fei and Liu, Zhaoyang and Wu, Shiwen and Gao, Jinyang and Zhang, Jiandong and Ding, Bolin and Cui, Bin},
  booktitle={2022 IEEE 38th international conference on data engineering (ICDE)},
  pages={1259--1273},
  year={2022},
  organization={IEEE}
}

@inproceedings{C2DSR,
  title={Contrastive cross-domain sequential recommendation},
  author={Cao, Jiangxia and Cong, Xin and Sheng, Jiawei and Liu, Tingwen and Wang, Bin},
  booktitle={Proceedings of the 31st ACM International Conference on Information \& Knowledge Management},
  pages={138--147},
  year={2022}
}

@article{TriCDR,
  title={Triple sequence learning for cross-domain recommendation},
  author={Ma, Haokai and Xie, Ruobing and Meng, Lei and Chen, Xin and Zhang, Xu and Lin, Leyu and Zhou, Jie},
  journal={ACM Transactions on Information Systems},
  volume={42},
  number={4},
  pages={1--29},
  year={2024},
  publisher={ACM New York, NY}
}

@inproceedings{LLM4CDSR,
  title={Bridge the Domains: Large Language Models Enhanced Cross-domain Sequential Recommendation},
  author={Liu, Qidong and Zhao, Xiangyu and Wang, Yejing and Zhang, Zijian and Zhong, Howard and Chen, Chong and Li, Xiang and Huang, Wei and Tian, Feng},
  booktitle={Proceedings of the 48th International ACM SIGIR Conference on Research and Development in Information Retrieval},
  pages={1582--1592},
  year={2025}
}

@inproceedings{tokenCDSR,
  title={RecGURU: Adversarial learning of generalized user representations for cross-domain recommendation},
  author={Li, Chenglin and Zhao, Mingjun and Zhang, Huanming and Yu, Chenyun and Cheng, Lei and Shu, Guoqiang and Kong, Beibei and Niu, Di},
  booktitle={Proceedings of the fifteenth ACM international conference on web search and data mining},
  pages={571--581},
  year={2022}
}

@inproceedings{sun_large_2024,
  title={Large language models enhanced collaborative filtering},
  author={Sun, Zhongxiang and Si, Zihua and Zang, Xiaoxue and Zheng, Kai and Song, Yang and Zhang, Xiao and Xu, Jun},
  booktitle={Proceedings of the 33rd ACM International Conference on Information and Knowledge Management},
  pages={2178--2188},
  year={2024}
}

@article{guo2025higarment,
  title={HiGarment: Cross-modal Harmony Based Diffusion Model for Flat Sketch to Realistic Garment Image},
  author={Guo, Junyi and Zhang, Jingxuan and Wu, Fangyu and Lu, Huanda and Wang, Qiufeng and Yang, Wenmian and Lim, Eng Gee and Lu, Dongming},
  journal={arXiv preprint arXiv:2505.23186},
  year={2025}
}

@inproceedings{zheng_lc-rec_2024,
  title={Adapting large language models by integrating collaborative semantics for recommendation},
  author={Zheng, Bowen and Hou, Yupeng and Lu, Hongyu and Chen, Yu and Zhao, Wayne Xin and Chen, Ming and Wen, Ji-Rong},
  booktitle={2024 IEEE 40th International Conference on Data Engineering (ICDE)},
  pages={1435--1448},
  year={2024},
  organization={IEEE}
}

@inproceedings{lin_bridging_2024-1,
  title={Bridging items and language: A transition paradigm for large language model-based recommendation},
  author={Lin, Xinyu and Wang, Wenjie and Li, Yongqi and Feng, Fuli and Ng, See-Kiong and Chua, Tat-Seng},
  booktitle={Proceedings of the 30th ACM SIGKDD Conference on Knowledge Discovery and Data Mining},
  pages={1816--1826},
  year={2024}
}

@inproceedings{bao_tallrec_2023,
	title = {{TALLRec}: {An} {Effective} and {Efficient} {Tuning} {Framework} to {Align} {Large} {Language} {Model} with {Recommendation}},
	shorttitle = {{TALLRec}},
	url = {http://arxiv.org/abs/2305.00447},
	doi = {10.1145/3604915.3608857},
	urldate = {2025-04-01},
	booktitle = {Proceedings of the 17th {ACM} {Conference} on {Recommender} {Systems}},
	author = {Bao, Keqin and Zhang, Jizhi and Zhang, Yang and Wang, Wenjie and Feng, Fuli and He, Xiangnan},
	month = sep,
	year = {2023},
	pages = {1007--1014},
}

@article{liu_enhancing_2025,
  title={Enhancing LLM-Based Recommendations Through Personalized Reasoning},
  author={Liu, Jiahao and Yan, Xueshuo and Li, Dongsheng and Zhang, Guangping and Gu, Hansu and Zhang, Peng and Lu, Tun and Shang, Li and Gu, Ning},
  journal={arXiv e-prints},
  pages={arXiv--2502},
  year={2025}
}

@article{zhang_adalora_2023,
  title={Adalora: Adaptive budget allocation for parameter-efficient fine-tuning},
  author={Zhang, Qingru and Chen, Minshuo and Bukharin, Alexander and Karampatziakis, Nikos and He, Pengcheng and Cheng, Yu and Chen, Weizhu and Zhao, Tuo},
  journal={arXiv preprint arXiv:2303.10512},
  year={2023}
}

@article{hu2022lora,
  title={Lora: Low-rank adaptation of large language models.},
  author={Hu, Edward J and Shen, Yelong and Wallis, Phillip and Allen-Zhu, Zeyuan and Li, Yuanzhi and Wang, Shean and Wang, Lu and Chen, Weizhu and others},
  journal={ICLR},
  volume={1},
  number={2},
  pages={3},
  year={2022}
}

@inproceedings{rqvae,
  title={Autoregressive image generation using residual quantization},
  author={Lee, Doyup and Kim, Chiheon and Kim, Saehoon and Cho, Minsu and Han, Wook-Shin},
  booktitle={Proceedings of the IEEE/CVF conference on computer vision and pattern recognition},
  pages={11523--11532},
  year={2022}
}

@article{quantiloss,
  title={Neural discrete representation learning},
  author={Van Den Oord, Aaron and Vinyals, Oriol and others},
  journal={Advances in neural information processing systems},
  volume={30},
  year={2017}
}

@inproceedings{amazon-dataset,
  title={Image-based recommendations on styles and substitutes},
  author={McAuley, Julian and Targett, Christopher and Shi, Qinfeng and Van Den Hengel, Anton},
  booktitle={Proceedings of the 38th international ACM SIGIR conference on research and development in information retrieval},
  pages={43--52},
  year={2015}
}

@inproceedings{douban1, title={A Graphical and Attentional Framework for Dual-Target Cross-Domain Recommendation}, author={Zhu, Feng and Wang, Yan and Chen, Chaochao and Liu, Guanfeng and Zheng, Xiaolin}, booktitle={Proceedings of the Twenty-Ninth International Joint Conference on Artificial Intelligence, IJCAI 2020}, pages={3001--3008}, year={2020} }

@inproceedings{douban2, title={DTCDR: A framework for dual-target cross-domain recommendation}, author={Zhu, Feng and Chen, Chaochao and Wang, Yan and Liu, Guanfeng and Zheng, Xiaolin}, booktitle={Proceedings of the 28th ACM International Conference on Information and Knowledge Management}, pages={1533--1542}, year={2019} }

@inproceedings{s3rec, series={CIKM ’20},
   title={S3-Rec: Self-Supervised Learning for Sequential Recommendation with Mutual Information Maximization},
   url={http://dx.doi.org/10.1145/3340531.3411954},
   DOI={10.1145/3340531.3411954},
   booktitle={Proceedings of the 29th ACM International Conference on Information \& Knowledge Management},
   publisher={ACM},
   author={Zhou, Kun and Wang, Hui and Zhao, Wayne Xin and Zhu, Yutao and Wang, Sirui and Zhang, Fuzheng and Wang, Zhongyuan and Wen, Ji-Rong},
   year={2020},
   month=oct, pages={1893–1902},
   collection={CIKM ’20} }

@article{leaveout,
author = {Zhao, Wayne Xin and Lin, Zihan and Feng, Zhichao and Wang, Pengfei and Wen, Ji-Rong},
title = {A Revisiting Study of Appropriate Offline Evaluation for Top-N Recommendation Algorithms},
year = {2022},
issue_date = {April 2023},
publisher = {Association for Computing Machinery},
address = {New York, NY, USA},
volume = {41},
number = {2},
issn = {1046-8188},
url = {https://doi.org/10.1145/3545796},
doi = {10.1145/3545796},
journal = {ACM Trans. Inf. Syst.},
month = dec,
articleno = {32},
numpages = {41}
}

@inproceedings{cross-domain-survey1,
  title={Cross-domain recommender systems: A survey of the state of the art},
  author={Fern{\'a}ndez-Tob{\'\i}as, Ignacio and Cantador, Iv{\'a}n and Kaminskas, Marius and Ricci, Francesco},
  booktitle={Spanish conference on information retrieval},
  volume={24},
  year={2012},
  organization={sn}
}

@inproceedings{llm-emb1,
  title={Llmemb: Large language model can be a good embedding generator for sequential recommendation},
  author={Liu, Qidong and Wu, Xian and Wang, Wanyu and Wang, Yejing and Zhu, Yuanshao and Zhao, Xiangyu and Tian, Feng and Zheng, Yefeng},
  booktitle={Proceedings of the AAAI Conference on Artificial Intelligence},
  volume={39},
  number={11},
  pages={12183--12191},
  year={2025}
}

@article{llm-survey1,
  title={A survey on large language models for recommendation},
  author={Wu, Likang and Zheng, Zhi and Qiu, Zhaopeng and Wang, Hao and Gu, Hongchao and Shen, Tingjia and Qin, Chuan and Zhu, Chen and Zhu, Hengshu and Liu, Qi and others},
  journal={World Wide Web},
  volume={27},
  number={5},
  pages={60},
  year={2024},
  publisher={Springer}
}

@article{li2024mixlora,
  title={Mixlora: Enhancing large language models fine-tuning with lora-based mixture of experts},
  author={Li, Dengchun and Ma, Yingzi and Wang, Naizheng and Ye, Zhengmao and Cheng, Zhiyuan and Tang, Yinghao and Zhang, Yan and Duan, Lei and Zuo, Jie and Yang, Cal and others},
  journal={arXiv preprint arXiv:2404.15159},
  year={2024}
}

@inproceedings{sun2019bert4rec,
  title={BERT4Rec: Sequential recommendation with bidirectional encoder representations from transformer},
  author={Sun, Fei and Liu, Jun and Wu, Jian and Pei, Changhua and Lin, Xiao and Ou, Wenwu and Jiang, Peng},
  booktitle={Proceedings of the 28th ACM international conference on information and knowledge management},
  pages={1441--1450},
  year={2019}
}

@inproceedings{fan2022stosa,
  title={Sequential recommendation via stochastic self-attention},
  author={Fan, Ziwei and Liu, Zhiwei and Wang, Yu and Wang, Alice and Nazari, Zahra and Zheng, Lei and Peng, Hao and Yu, Philip S},
  booktitle={Proceedings of the ACM web conference 2022},
  pages={2036--2047},
  year={2022}
}

@inproceedings{vqrec,
  title={Learning vector-quantized item representation for transferable sequential recommenders},
  author={Hou, Yupeng and He, Zhankui and McAuley, Julian and Zhao, Wayne Xin},
  booktitle={Proceedings of the ACM Web Conference 2023},
  pages={1162--1171},
  year={2023}
}

@article{hstu,
  title={Actions speak louder than words: Trillion-parameter sequential transducers for generative recommendations},
  author={Zhai, Jiaqi and Liao, Lucy and Liu, Xing and Wang, Yueming and Li, Rui and Cao, Xuan and Gao, Leon and Gong, Zhaojie and Gu, Fangda and He, Michael and others},
  journal={arXiv preprint arXiv:2402.17152},
  year={2024}
}

@article{alemi2016deep,
  title={Deep variational information bottleneck},
  author={Alemi, Alexander A and Fischer, Ian and Dillon, Joshua V and Murphy, Kevin},
  journal={arXiv preprint arXiv:1612.00410},
  year={2016}
}

@article{liu2024fine,
  title={Fine Tuning Out-of-Vocabulary Item Recommendation with User Sequence Imagination},
  author={Liu, Ruochen and Chen, Hao and Bei, Yuanchen and Shen, Qijie and Zhong, Fangwei and Wang, Senzhang and Wang, Jianxin},
  journal={Advances in Neural Information Processing Systems},
  volume={37},
  pages={8930--8955},
  year={2024}
}

@inproceedings{yin2025unleash,
  title={Unleash LLMs Potential for Sequential Recommendation by Coordinating Dual Dynamic Index Mechanism},
  author={Yin, Jun and Zeng, Zhengxin and Li, Mingzheng and Yan, Hao and Li, Chaozhuo and Han, Weihao and Zhang, Jianjin and Liu, Ruochen and Sun, Hao and Deng, Weiwei and others},
  booktitle={Proceedings of the ACM on Web Conference 2025},
  pages={216--227},
  year={2025}
}

@article{liu2024unihr,
  title={UniHR: Hierarchical Representation Learning for Unified Knowledge Graph Link Prediction},
  author={Liu, Zhiqiang and Hua, Yin and Chen, Mingyang and Zhang, Yichi and Chen, Zhuo and Liang, Lei and Chen, Huajun and Zhang, Wen},
  journal={arXiv preprint arXiv:2411.07019},
  year={2024}
}

@article{zhang2025mitigating,
  title={Mitigating propensity bias of large language models for recommender systems},
  author={Zhang, Guixian and Yuan, Guan and Cheng, Debo and Liu, Lin and Li, Jiuyong and Zhang, Shichao},
  journal={ACM Transactions on Information Systems},
  volume={43},
  number={6},
  pages={1--26},
  year={2025},
  publisher={ACM New York, NY}
}

@article{li2025multi,
  title={Multi-Objective Unlearning in Recommender Systems via Preference Guided Pareto Exploration},
  author={Li, Yuyuan and Zhang, Yizhao and Liu, Weiming and Feng, Xiaohua and Han, Zhongxuan and Chen, Chaochao and Yan, Chenggang},
  journal={IEEE Transactions on Services Computing},
  year={2025},
  publisher={IEEE}
}

@article{chen2024post,
  title={Post-training attribute unlearning in recommender systems},
  author={Chen, Chaochao and Zhang, Yizhao and Li, Yuyuan and Wang, Jun and Qi, Lianyong and Xu, Xiaolong and Zheng, Xiaolin and Yin, Jianwei},
  journal={ACM Transactions on Information Systems},
  volume={43},
  number={1},
  pages={1--28},
  year={2024},
  publisher={ACM New York, NY, USA}
}

@inproceedings{lu2025dmmd4sr,
  title={DMMD4SR: Diffusion Model-based Multi-level Multimodal Denoising for Sequential Recommendation},
  author={Lu, Weihai and Yin, Li},
  booktitle={Proceedings of the 33rd ACM International Conference on Multimedia},
  pages={6363--6372},
  year={2025}
}

@inproceedings{cui2025multi,
  title={Multi-Modal Multi-Behavior Sequential Recommendation with Conditional Diffusion-Based Feature Denoising},
  author={Cui, Xiaoxi and Lu, Weihai and Tong, Yu and Li, Yiheng and Zhao, Zhejun},
  booktitle={Proceedings of the 48th International ACM SIGIR Conference on Research and Development in Information Retrieval},
  pages={1593--1602},
  year={2025}
}

@inproceedings{mo2024min,
  title={MIN: Multi-stage Interactive Network for Multimodal Recommendation},
  author={Mo, Minghao and Lu, Weihai and Xie, Qixiao and Lv, Xiang and Xiao, Zikai and Yang, Hong and Zhang, Yanchun},
  booktitle={International Conference on Web Information Systems Engineering},
  pages={191--205},
  year={2024},
  organization={Springer}
}

@inproceedings{li2024mhhcr,
  title={MHHCR: Multi-behavior Heterogeneous Hypergraph Contrastive Recommendation},
  author={Li, Yiheng and Lu, Weihai},
  booktitle={International Conference on Web Information Systems Engineering},
  pages={91--102},
  year={2024},
  organization={Springer}
}

@inproceedings{zeng2025numina,
  title     = "{NUMINA}: A Natural Understanding Benchmark for Multi-dimensional Intelligence and Numerical Reasoning",
  author    = "Zeng, Changyu and Wang, Yifan and Wang, Zimu and Wang, Wei and Yang, Zhengni and Bao, Muyi and Xiao, Jimin and Nguyen, Anh and Yue, Yutao",
  booktitle = "Findings of the Association for Computational Linguistics: EMNLP 2025",
  year      = "2025",
  pages     = "22575--22590"
}

@inproceedings{xiang2025harnessing,
  author    = {Yang Xiang and Li Fan and Chenke Yin and Menglin Kong and Chengtao Ji},
  title     = {Harnessing Light for Cold-Start Recommendations via Epistemic Uncertainty},
  booktitle = {Proceedings of the 34th ACM International Conference on Information and Knowledge Management (CIKM)},
  year      = {2025},
  pages     = {5361--5365}
}

@inproceedings{zhang2025llm,
  title={LLM-Driven Completeness and Consistency Evaluation for Cultural Heritage Data Augmentation in Cross-Modal Retrieval},
  author={Zhang, Jian and Guo, Junyi and Yuan, Junyi and Lu, Huanda and Zhou, Yanlin and Wu, Fangyu and Wang, Qiufeng and Lu, Dongming},
  booktitle={Proceedings of the 2025 Conference on Empirical Methods in Natural Language Processing},
  pages={19418--19428},
  year={2025}
}

@inproceedings{zhang2024goal,
  title={Goal-guided generative prompt injection attack on large language models},
  author={Zhang, Chong and Jin, Mingyu and Yu, Qinkai and Liu, Chengzhi and Xue, Haochen and Jin, Xiaobo},
  booktitle={2024 IEEE International Conference on Data Mining (ICDM)},
  pages={941--946},
  year={2024},
  organization={IEEE}
}

@article{mo2025one,
  title={One multimodal plugin enhancing all: CLIP-based pre-training framework enhancing multimodal item representations in recommendation systems},
  author={Mo, Minghao and Lu, Weihai and Xie, Qixiao and Xiao, Zikai and Lv, Xiang and Yang, Hong and Zhang, Yanchun},
  journal={Neurocomputing},
  volume={637},
  pages={130059},
  year={2025},
  publisher={Elsevier}
}

@inproceedings{li2023making,
  title={Making users indistinguishable: Attribute-wise unlearning in recommender systems},
  author={Li, Yuyuan and Chen, Chaochao and Zheng, Xiaolin and Zhang, Yizhao and Han, Zhongxuan and Meng, Dan and Wang, Jun},
  booktitle={Proceedings of the 31st ACM International Conference on Multimedia},
  pages={984--994},
  year={2023}
}

@article{li2024survey,
  title={A survey on recommendation unlearning: Fundamentals, taxonomy, evaluation, and open questions},
  author={Li, Yuyuan and Feng, Xiaohua and Chen, Chaochao and Yang, Qiang},
  journal={arXiv preprint arXiv:2412.12836},
  year={2024}
}

@inproceedings{yuan2025towards,
  title={Towards Cross-Modal Retrieval in Chinese Cultural Heritage Documents: Dataset and Solution},
  author={Yuan, Junyi and Zhang, Jian and Wu, Fangyu and Lu, Huanda and Lu, Dongming and Wang, Qiufeng},
  booktitle={International Conference on Document Analysis and Recognition},
  pages={570--586},
  year={2025},
  organization={Springer}
}

\end{document}